\documentclass[preprint2]{aastex61}
\newcommand{\src}{PSR~J1119$-$6127}
\newcommand{\source}{PSR~J1119$-$6127}

\shorttitle{Magnetar-like X-ray bursts of \source{}}
\shortauthors{Archibald et al.}

\usepackage{lineno, soul, color}

\begin{document}

\title{Magnetar-like X-ray bursts suppress pulsar radio emission}

\correspondingauthor{R. F. Archibald }
\email{archibald@astro.utoronto.ca}

\author{R. F. Archibald}
\affiliation{Department of Physics \& McGill Space Institute, McGill University,
	         3600 University Street, Montreal QC, H3A 2T8, Canada}
\affiliation{Department of Astronomy and Astrophysics, University of Toronto, 50 St. George Street, Toronto, ON M5S 3H4, Canada}

\author{M. Burgay}
\affiliation{INAF Osservatorio Astronomico di Cagliari, Via della Scienza 5, I-09047 Selargius, Italy}

\author{M.~Lyutikov}
\affiliation{Department of Physics, Purdue University, 525 Northwestern Avenue, West Lafayette, IN 47907-2036, USA}
\affiliation{Department of Physics \& McGill Space Institute, McGill University,
	3600 University Street, Montreal QC, H3A 2T8, Canada}
\author{V.~M.~Kaspi}
\affiliation{Department of Physics \& McGill Space Institute, McGill University,
	3600 University Street, Montreal QC, H3A 2T8, Canada}
\author{P.~Esposito}
\affiliation{Anton Pannekoek Institute for Astronomy, University of Amsterdam, Postbus 94249, 1090\,GE Amsterdam, The Netherlands}
\author{G.~Israel}
\affiliation{INAF – Osservatorio Astronomico di Roma, via Frascati 33, I-00040 Monteporzio Catone, Roma, Italy}
\author{M.~Kerr}
\affiliation{Space Science Division, Naval Research Laboratory, Washington, DC 20375-5352, USA}
\author{A.~Possenti}
\affiliation{INAF Osservatorio Astronomico di Cagliari, Via della Scienza 5, I-09047 Selargius, Italy}
\author{N.~Rea}
\affiliation{Anton Pannekoek Institute for Astronomy, University of Amsterdam, Postbus 94249, 1090\,GE Amsterdam, The Netherlands}
\affiliation{Institute of Space Sciences (ICE, CSIC–IEEC), Carrer de Can Magrans, S/N, 08193, Barcelona, Spain}
\author{J.~Sarkissian}
\affiliation{CSIRO Astronomy and Space Science, Parkes Observatory, PO Box 276, Parkes NSW 2870, Australia}
\author{P.~Scholz}
\affiliation{National Research Council of Canada, Herzberg Astronomy and Astrophysics, Dominion Radio Astrophysical Observatory, P.O. Box 248, Penticton, BC V2A 6J9, Canada}
\author{S.~P.~Tendulkar}
\affiliation{Department of Physics \& McGill Space Institute, McGill University,
	3600 University Street, Montreal QC, H3A 2T8, Canada}


\begin{abstract}
Rotation-powered pulsars and magnetars are two different observational manifestations of neutron stars: rotation powered pulsars are rapidly spinning objects that are mostly observed as pulsating radio sources, while magnetars, neutron stars with the highest known magnetic fields, often emit short-duration X-ray bursts. Here we report simultaneous observations of the high-magnetic-field radio pulsar \src{} at X-ray, with {\it XMM-Newton} \& {\it NuSTAR}, and at radio energies with Parkes radio telescope, during a period of magnetar-like bursts.
The rotationally powered radio emission shuts off coincident with the occurrence of multiple X-ray bursts, and recovers on a time scale of $\sim$ 70 seconds. These observations of related radio and X-ray phenomena further solidify the connection between radio pulsars and magnetars, and suggest that the pair plasma produced in bursts can disrupt the acceleration mechanism of radio emitting particles.  
\end{abstract}

\section{Introduction}
\src{} is one of the youngest known pulsars (characteristic age of $\sim1600$\,yr) and has a magnetic field of $B$\,=\,$4\times10^{13}$\,G, one of the highest known among radio pulsars \citep{2000ApJ...541..367C}, where $B= 3.2 \times 10^{19} \sqrt{P \dot{P}}$\,G is the inferred equatorial dipolar field strength, given the pulsar's rotation period $P = 0.4$~s and its spin-down rate $\dot{P} = 4.0\times10^{-12}$.
While generally observed as a stable radio pulsar, in 2007 it exhibited an unusual spin-up glitch accompanied by a short-lived change in its radio pulse profile \citep{2011MNRAS.411.1917W}.
Prior to this event, only magnetars, and a lone rotation powered pulsar, had exhibited rotational glitches with any radiative signature \citep{2008APJ.673..1044D, 2008Sci...319.1802G}.

\begin{figure}[!t]
	\includegraphics[width=\columnwidth]{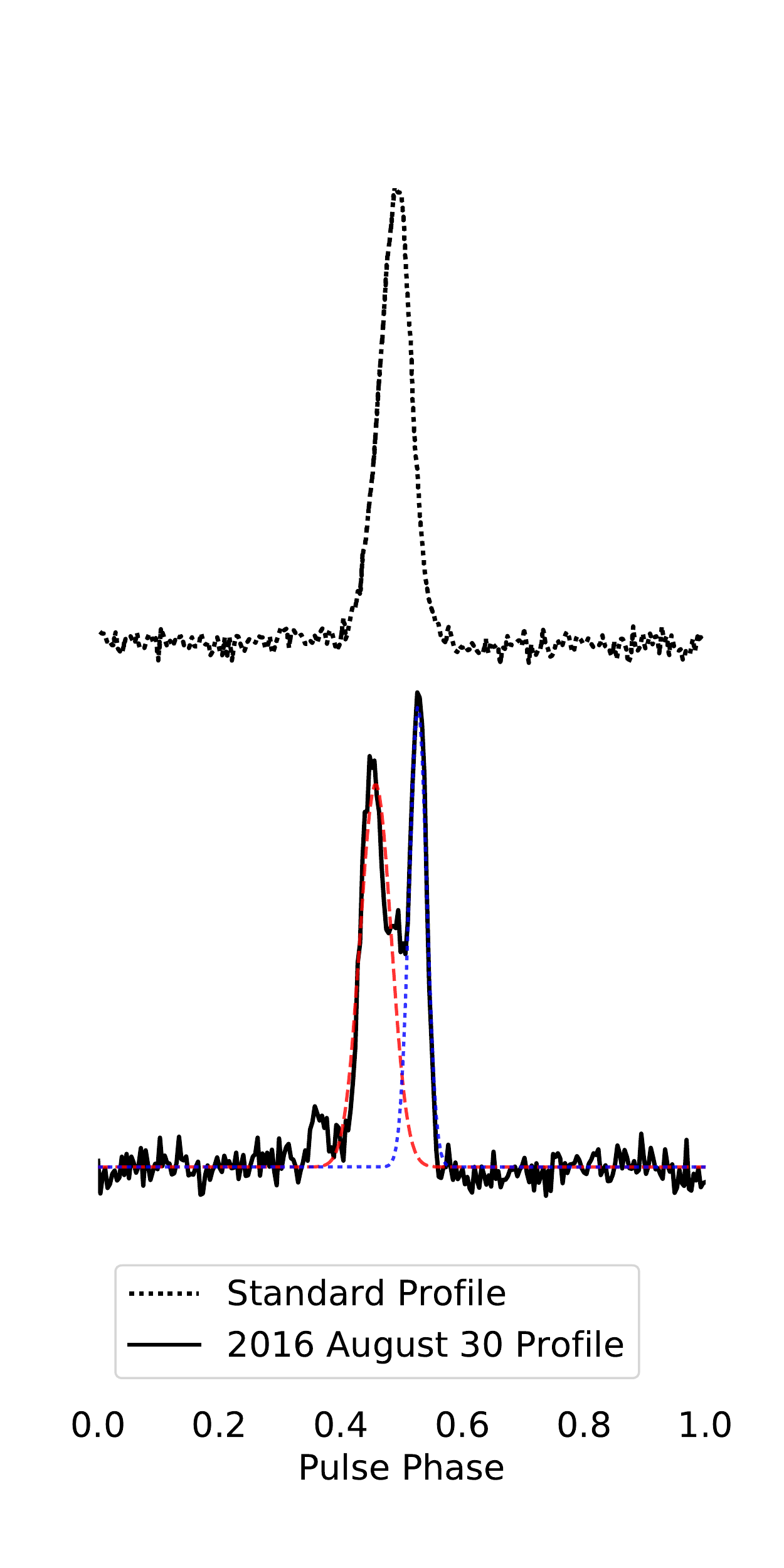}
	\caption{1.4\,GHz radio pulse profiles of \src{}. The dashed line shows the standard pulse profile \citep{2000ApJ...541..367C}, and the solid line displays the profile during the 30 August 2016 Parkes observation. The red and blue overlays denote the Gaussian fits to peaks one and two respectively.}
	\label{fig:prof}
\end{figure}

On 27 July 2016, \src{} exhibited a flux increase by a factor of $>160$ in the 0.5--10\,keV band, emitted several short magnetar-like bursts, and had a rotational glitch \citep{2016ApJ...829L..21A, 2016ApJ...829L..25G}. 
At this time \src{} also changed its radio emission characteristics, initially turning off as a radio pulsar,  before returning and displaying a magnetar-like radio spectral flattening \citep{2016ATel.9870....1P}.
As well, after the radio re-activation, the normal emission profile, well modeled as a single Gaussian peak \citep{2000ApJ...541..367C}, changed to a two-peaked pulse profile \citep{2017ApJ...834L...2M}, similar to that seen following the 2007 glitch, see Figure~\ref{fig:prof}.

Here, we present the results from  simultaneous observations of magnetar-like bursts from \src{} with the Parkes 64-m radio telescope at an observing frequency of 1369\,MHz, and in the X-ray band using {\it XMM-Newton} \citep{2001A&A...365L...1J}, and {\it NuSTAR} \citep{2013ApJ...770..103H} taken on 30 August  2016 (MJD 57630), 33 days after the initial outburst activation.

\phantom{\\}

\section{Observations \& Analysis}
\subsection{Parkes Radio Telescope}
\src{} was observed on  2016 August 30 at the Parkes 64-m radio telescope, both as a part of the NAPA\footnote{Non-A-Priori-Assignable, see\\ \texttt{http://www.atnf.csiro.au/observers/apply/\\\phantom{http://}too\_apply.html\#Non\%20A-priori\%Assignable}.}
propopal P626 (P.I. Burgay) and as part of the regular timing program P574 (P.I. Kerr). 
The observations were carried out with the H--OH receiver over a bandwidth of 256\,MHz centered at a frequency of 1369\,MHz and split in 1024 frequency channels. 
Data were collected with the ATNF DFB4 digital backend in search mode, 4-bit-sampling the signal every 256\,$\mu$s.
Pulse phases were assigned to each data-sample using a local pulsar ephemeris.

The radio data reduction was performed with
{\ttfamily{PRESTO}}\footnote{ \texttt{http://www.cv.nrao.edu/$\sim$sransom/presto/}.} \citep{2002AJ....124.1788R}.
Data were first cleaned of the most prominent radio frequency interference using {\ttfamily{rfifind}}, corrected for the observed dispersion measure of 706.5(3)\,pc\,cm$^{-3}$, and corrected to the Solar System barycenter using {\ttfamily{prepdata}}.

The radio fluence, the integrated flux under the pulse profile, was measured by fitting the pulse profile as the sum of two Gaussian components using a least-squares minimization. 

\subsection{ {\it XMM-Newton} \& {\it NuSTAR}}

The {\it XMM-Newton} observation analyzed here (ObsID 0741732801)  was performed with both the EPIC/pn and EPIC/MOS cameras in Small Window mode.
These modes provide 5.7-ms and 0.3-s time resolution, respectively.
{\it XMM} Science Analysis System (SAS) version 16.0 and {\tt HEASOFT v6.19} were used to reduce the data.
The raw Observation Data Files (ODF) were first downloaded for each observation and were then pre-processed using the SAS tools {\tt emproc} and {\tt epproc}.
The events were filtered so that single--quadruple events with energies between 0.1--12\,keV (pn) and 0.2--15\,keV (MOS) were retained, and standard ``FLAG'' filtering was applied.
The light curves were then inspected for soft proton flares, however, there were none detected in the overlapping time.  
In this work, since we were searching for X-ray bursts, we used only the high-time-resolution pn data.
We extracted source events from an 18'' radius region centered on \src{}.

The {\it NuSTAR} observation (ObsID 80102048008) was reduced using the {\tt nupipeline} scripts, using {\tt HEASOFT v6.20}.
Source events were extracted within a 30 pixel (72'') radius around the centroid. Appropriate background regions were selected from the same detector as the source location. Spectra were extracted using the {\tt nuproducts} script.
Using {\tt grppha} channels 0--35 ($<3$ keV) and 1935--4095 ($> 79$ keV) were ignored, and all good channels were binned to have a minimum of one count per energy bin.

All X-ray data were then corrected so that their arrival times were referenced to the Solar System barycenter using the {\it Chandra} location of \src{} \citep{2003ApJ...591L.143G}.

\section{Magnetar-like X-ray Bursts Characteristics}
At the epoch of the X-ray observations reported here, the persistent 0.5--10\,keV absorbed X-ray flux was $6.22(6)\times10^{-12}$\,erg\,s$^{-1}$\,cm$^{-2}$, still a factor of $\sim50$ higher than its quiescent flux, but a factor of $\sim 6$ lower than that measured at the outburst peak \citep{2016ApJ...829L..21A}.
The X-ray flux was modulated at the spin-period with a root-mean-squared pulse fraction in this band of 52(2)\%.

\begin{figure}[]
	\includegraphics[width=\columnwidth]{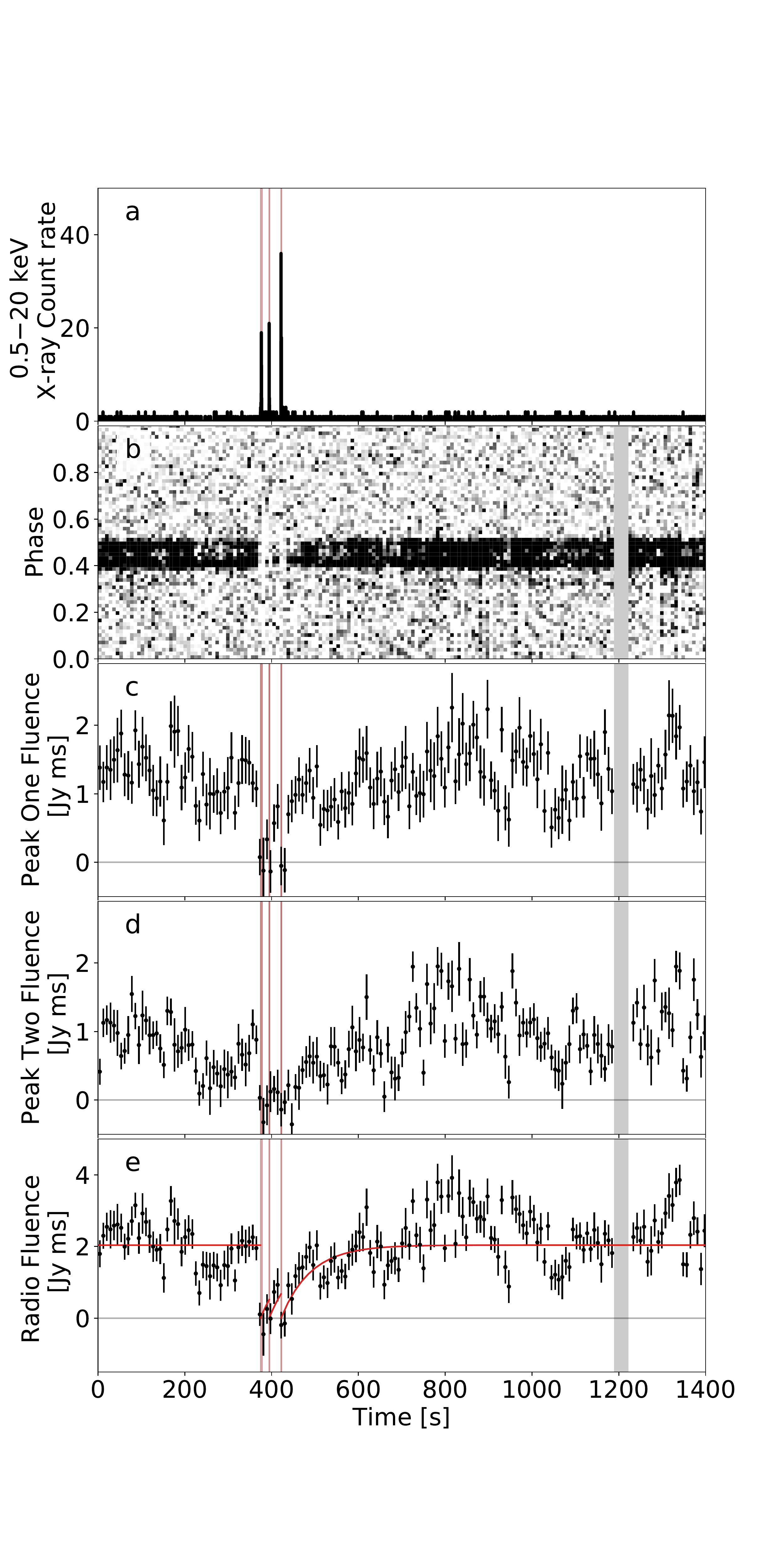}
	\caption{X-ray and 1.4\,GHz radio observations of \src{}.
		{\bf(a)} The combined {\it XMM} and {\it NuSTAR} 0.5--20\,keV count rate over time.
		{\bf (b)} The gray-scale shows the 1.4\,GHz radio intensity as a function of pulse phase and time. The plotted fluences are averages over 20 pulse periods. 
		{\bf (c)} Time evolution of the measured pulsed radio fluence for the first peak.
		{\bf (d)}  Time evolution of the measured pulsed radio fluence for the second peak.
		{\bf(e) }The total pulsed radio fluence over time.
		The red line shows the best-fit exponential recovery model with a time scale of 70(10)\,s.
		The grayed out region at 1200\,s is due to a radio frequency interference spike. Note that the time delay due to dispersion has been corrected for the observed dispersion measure of 706.5(3)\,pc\,cm$^{-3}$. The red areas indicate the $T_{90}$ of the three X-ray bursts. }
	\label{fig:portrait}
\end{figure}

During these observations, we detected three magnetar-like X-ray bursts in both the {\it XMM} and {\it NuSTAR} data, see Figure~\ref{fig:portrait}, panel a.
Each burst has a total energies of $\sim 10^{37}$\,erg in the 0.5--70\,keV band and is well-modeled by a blackbody spectrum.
In Table~\ref{tab:bursts} we quantify the burst properties: the time of the burst peak, the spectral properties, and $T_{90}$ using the method described in \cite{2004ApJ...607..959G}.

\begin{figure}[]
	\includegraphics[width=\columnwidth]{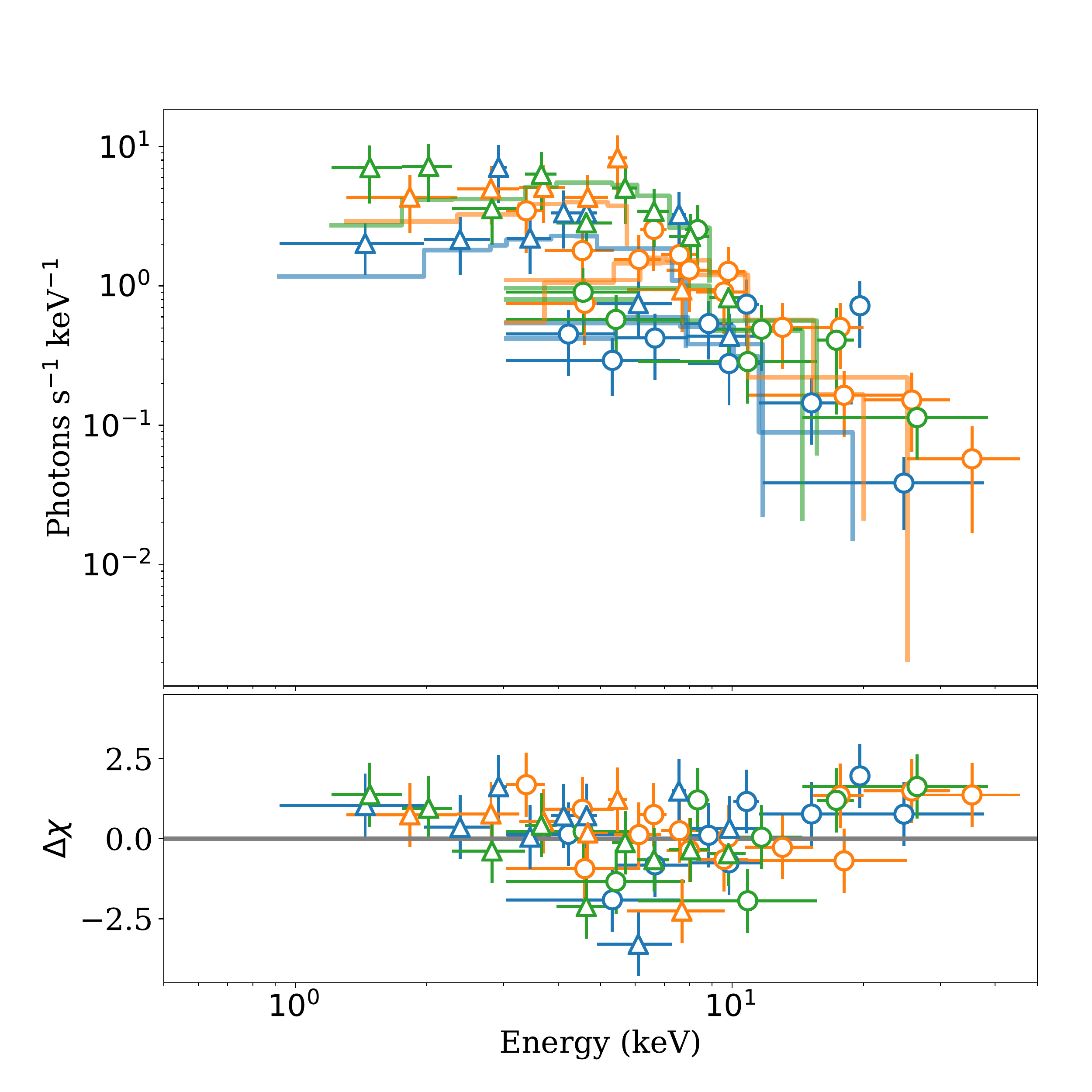}
	\caption{X-ray spectra and residuals of the three bursts. Data from {\it XMM} are plotted as triangles, and {\it NuSTAR} as circles. The three bursts are plotted in different colors with the first being blue, the second orange, and the third green. The lines indicate the model presented in Table~\ref{tab:bursts}, and the bottom plot shows the residuals of the data to the model in units of $\Delta\chi$. }
	\label{fig:spec}
\end{figure}

To fit the burst spectra, events occurring within the $T_{90}$ of each burst were extracted.
Using {\tt xspec v12.9.1}, the spectra  were fit to an absorbed blackbody using Cash statistics \citep{1979ApJ...228..939C} for fitting and parameter estimation of the unbinned data. $N_H$ was held fixed to $1.2\times10^{22}$\,cm$^2$\citep{2016ApJ...829L..21A} using {\tt wilm} abundances and {\tt vern} photoelectric cross-sections.
The spectra of the three bursts, as well as the residuals to that fit are displayed in Figure~\ref{fig:spec}.


\begin{table*}
	\begin{center}
		\caption{X-ray Burst Properties.}
		\label{tab:bursts}
		\begin{tabular}{ccccccc}
			\hline
			\hline
			Burst time &T$_{90}^\dagger$ & kT  & 0.5--70\,keV Total Energy in T$_{90} ^*$  & C-stat/dof$^\ddagger$\\ 
			MJD  &ms & keV & 10$^{36}$ erg \\
			57630.1914063 & 4100(300) &$2.5_{-0.4}^{+0.6}$ &$10._{-2}^{+3}$ &120.1/76\\
			57630.1916137 & 1600(150) &$2.6_{-0.6}^{+0.8}$ &$7._{-2}^{+3}$ &66.2/51\\
			57630.1919302 & 1900(300) &$3.0_{-0.5}^{+0.7}$ &$14._{-3}^{+4}$&72.9/87\\
			\hline
			\hline
			\newline
		\end{tabular}
		\newline
		$^\dagger$ The duration in which the burst emitted 90\% of its fluence.\\
		$^*$ Assuming isotropic emission at a distance of 8.4\,kpc \citep{2004MNRAS.352.1405C}.\\
		$^\ddagger$ Cash-statistic \citep{1979ApJ...228..939C} compared to the number of degrees of freedom for the spectral fit.
	\end{center}
	
\end{table*}

While these bursts would be unremarkable from a typical magnetar \citep{2015ApJS..218...11C, 2015ApJ...807...93A}, interestingly in this case the bursts last longer than the rotation period of the pulsar. 

\begin{figure}[]
	\includegraphics[width=\columnwidth]{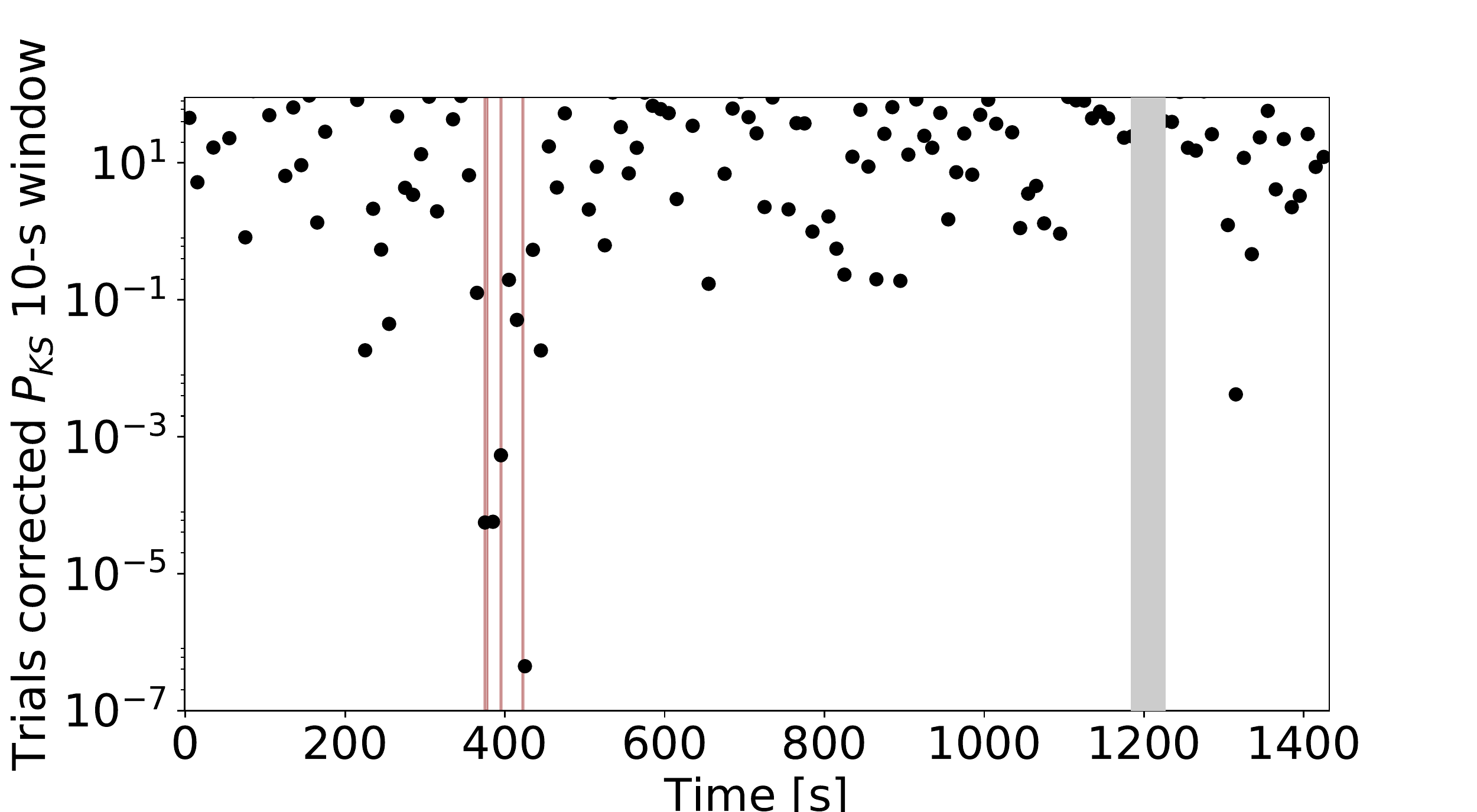}
	\caption{Trials-corrected two-sample Kolmogorov-Smirnov test on the single pulse fluences comparing each 10-s of data to all other observed data to determine the probability of rejecting the null hypothesis that two samples are drawn from the same parent population.
		The grayed out region at 1200\,s is due to a radio frequency interference spike. Note that the time delay due to dispersion has been corrected for the observed dispersion measure of 706.5(3)\,pc\,cm$^{-3}$. The red areas indicate the $T_{90}$ of the three X-ray bursts.  }
	\label{fig:KStest}
\end{figure}

Coincident to within 5 pulse periods ($\sim2$\,s) with the first of these bursts, the pulsed radio fluence at 1369\,MHz dropped from a mean of 2.04(5)\,Jy\,ms to undetected, with a 99\% confidence upper limit of 0.5\,Jy\,ms for the interval between the first and second bursts (Figure~\ref{fig:portrait}).
The pulsed radio fluence also drops to be consistent with zero at the times of the second and third burst.
The disappearance of the radio emission coincident with the X-ray bursts is clear in the gray-scale (panel b).
Although the radio pulse fluences vary significantly, we verified that the coincidence of the radio disappearance with the X-ray bursts is not due to chance as follows.
We divided the radio fluence time series into 10-s intervals and compared the single-pulse fluence distribution in that 10-s interval to that of the remainder of the observation.
This was done by means of a two-sample Kolmogorov-Smirnov (KS) test.
For the times surrounding the bursts, the null hypothesis that the two samples were drawn from the same underlying distribution can be rejected with a  probabilities ranging from $P = 10^{-4}- 3\times 10^{-7}$, compared to a minimum probability of $P = 1\times 10^{-2}$ outside the burst region, see Figure~\ref{fig:KStest}.
Thus we can say with high confidence that the radio fluences surrounding the X-ray bursts  significantly drops at the time of the X-ray bursts.

The second radio peak became undetectable for the  95(5)\,s following the first burst (Figure~\ref{fig:portrait}).
The radio fluence of the first peak behaved more  sporadically.
It too became undetectable at the time of the first X-ray burst, but increased to become detectable between bursts two and three, reaching a a peak fluence of  0.8(3)\,Jy\,ms before again becoming undetectable at the onset of the third burst.

Based on the apparently similar time scale of recovery after the second and third burst, we co-fit the radio recovery following each burst with a single time scale.
The total radio fluence following each burst can be modeled as an exponential recovery, with a time scale of 70(10)\,s.

Since the probability for chance coincidence of the radio shut-off with the X-ray bursts is negligibly small, we  next consider possible physical mechanisms for this suppression that could arise in the magnetosphere of the neutron star.

\section{Discussion }
The structures of pulsar and magnetar magnetospheres are complicated.
In radio pulsars the magnetosphere is separated into open and closed magnetic field zones, depending on whether a given field line returns to the star or extends to infinity.
In standard pulsar models, the magnetosphere near the stellar surface is populated
with plasma having the Goldreich-Julian density  \citep{1969ApJ...157..869G}
\begin{equation} 
\rho_{GJ} = - \frac{1}{2\pi c e} \mathbf{\Omega} \cdot \mathbf{B} \simeq 7\times10^{12} \mathrm{pairs\,per\,cm}^{3}
\end{equation}  
where  $\mathbf{\Omega}$ is the angular velocity of the pulsar, $c$ is the speed of light, $e$ is the charge of the electron, and $\mathbf{B}$ is the surface magnetic field. 
In the special regions on the open field lines, called the ``gaps'', where the local density is below the Goldreich-Julian density, electric fields parallel to the magnetic field accelerate particles to ultra-relativistic energies that initiate an electron-positron pair cascade.
The development of plasma instabilities is thought to lead to the production of coherent radio emission \citep{1995JApA...16..137M}. It is important that the gaps have relatively small plasma density - otherwise the accelerating electric field is suppressed by the charge separation.

In the case of magnetars, the dissipation of magnetic energy in the  magnetosphere 
creates a  trapped pair-plasma fireball \citep{1995MNRAS.275..255T}. 
For a distance of 8.4 kpc, a flux of $\sim10^{-9}$\,erg$\,$s$^{-1}$\,cm$^{2}$ and the  blackbody temperature of  2\,keV, the effective radius of the trapped pair fireball is $R_{fb} \approx 1$\,km. Since the total radiated energy is $E_{rad} \sim 10^{37}$\,erg, the magnetic field needed to confine the fireball is 
$B \geq \sqrt{6 E_{rad}}/R_{fb}^{3/2} = 2.5 \times 10^{11}$\,G. For the surface magnetic field of $4 \times  10^{13}$\,G the fireball can be trapped at radii somewhat larger than the stellar radius, $\leq 6 R_{NS}$. 

The core of a pair-plasma fireball has a typical temperature of $ k_B T \sim m_e c^2$, where $m_e$ is the mass of an electron \citep{1995MNRAS.275..255T}.
If more energy is added to the pair fireball it is used mostly to create electron-positron pairs, not to increased thermal motion.
Given the total radiated/dissipated energy $E_{rad}$ and the estimates of the size of the fireball $R_{fb}$, the typical pair density is $n_\pm \sim E_{rad}/(m_e c^2) (4 \pi/3) R_{fb}^{3} = 3 \times 10^{27} $ particles per cubic centimeter. This exceeds the Goldreich-Julian density by 15 orders of magnitude.
Thus, if even a small fraction of the created pairs ``leak'' from the trapped fireball into the gap region of acceleration of radio-emitting particles, the acceleration will be suppressed and the radio emission switched off.
We believe that the evaporation of this fireball sets the shut-off timescale for the radio emission.
We note that this should also cause a temporary cessation of gamma-ray and magnetospheric X-ray pulsations.
This, however, is not detectable with current telescopes as the X-ray emission in \src{} is mainly thermal, and a gamma-ray detection takes weeks of integration  \citep{2016ApJ...829L..21A}.

Alternative explanations  for the observed anti-correlation between the magnetar-like bursts and radio emission could be related to  the radical changes or even the total disappearance of radio emission routinely seen in mode-changing and nulling radio pulsars.
Indeed, simultaneous X-ray state changes \citep{2013Sci...339..436H, 2016ApJ...831...21M} have been observed in some nullers.
Nulling and mode-changes are likely to be non-linear magnetospheric phenomena whereby the overall structure switches between different states, with different global current structures and emission properties.
These global changes can be driven either by variations in the position of the edge of the magnetosphere \citep{2010MNRAS.408L..41T} or due to ``twisting'' of the NS’s magnetic field lines by crustal motions  \citep{2012ApJ...754L..12P,2016ApJ...827...80H}.
However, the time scale predicted for the changes in the twist is on the order of days -- not the $\sim$100\,s we observe in \src{}.
In either case, a global change in the magnetosphere is expected to be accompanied by correlated changes in $\dot{P}$ and profile, as seen in some nullers and mode changing pulsars \citep{2006Sci...312..549K, 2010Sci...329..408L}. 
We are insensitive to a change in the spin-down rate of \src{} over the short time span of the nulls. 

This simultaneous radio and X-ray observation of  magnetar-like bursts may also have relevance to fast radio bursts (FRBs).
FRBs are bright, few-ms duration radio bursts from a yet unknown extragalactic origin \citep{2007Sci...318..777L, 2013Sci...341...53T}.
With the discovery of a repeating FRB \citep{2016Natur.531..202S}, and a confirmation of its extragalactic origin \citep{2017Natur.541...58C, 2017ApJ...834L...7T}, a magnetar origin for FRBs has become a favored model \citep{2017ApJ...838L..13L,2017ApJ...843...84N}.  
The non-detection of a bright radio counterpart to the bursts observed from \src{}, especially when taken with the radio non-detection of the SGR\,1806$-$20 giant flare \citep{2016ApJ...827...59T}, impose a constraint on a magnetar origin for FRBs.
We do however caution, in the case of \src{}, we observed a shut off of the rotationally powered radio emission, and radio emission from FRBs may be magnetically powered  \citep{2017ApJ...838L..13L}.

\acknowledgements
We thank an anonymous referee for a thorough reading which improved the quality of the manuscript.
This work made use of data from the {\it NuSTAR} mission, a project led by the California Institute of Technology, managed by the Jet Propulsion Laboratory, and funded by the National Aeronautics and Space Administration.
This work made use of observations obtained with XMM-Newton, an ESA science mission with instruments and contributions directly funded by ESA Member States and NASA.
Parkes radio telescope is part of the Australia Telescope National Facility which is funded by the Commonwealth of Australia for operation as a National Facility managed by CSIRO.
R.F.A. acknowledges support from an NSERC  Alexander Graham Bell Canada Graduate Scholarship.
M.L. is supported by NSF grant AST-1306672 and DoE grant DE-SC0016369.
V.M.K. receives support from an NSERC Discovery Grant and Accelerator Supplement, Centre de Recherche en Astrophysique du Queb\'ec, an R. Howard Webster Foundation Fellowship from the Canadian Institute for Advanced Study, the Canada Research Chairs Program, and the Lorne Trottier Chair in Astrophysics and Cosmology.
P.E. and N.R. acknowledge funding in the framework of the NWO Vidi award A.2320.0076.
P.S. holds a Covington Fellowship at DRAO.
Work at NRL is supported by NASA.


\facilities{{\it NuSTAR, XMM-Newton}, Parkes Radio Telescope}
	


\bibliographystyle{yahapj}

\end{document}